# (Legal Design) Research through Litigation

Reuben Kirkham, Monash University, Australia

reuben@onlinecourt.uk

This paper proposes the concept of 'research through litigation', where a HCI researcher would bring a claim in the legal system in order to understand judicial attitudes towards technologies. Based on my seven years of experience of bringing legal cases as a computer scientist in Tribunals, I demonstrate the value of this approach by presenting multiple case studies, which illustrate the counter-intuitive approach towards technology taken by Tribunals. This exercise surfaced some serious (and somewhat surreal) concerns with the operation of the justice system, as well as demonstrating how research through litigation changed the law on several occasions. This work therefore makes important methodological and practical contributions to the nascent topic of legal (interaction) design, especially from a methodological standpoint.

## 1 INTRODUCTION

HCI researchers have often conducted fieldwork in challenging settings, sometimes including settings involving vulnerable populations that raise serious ethical concerns (e.g. [5,7,20]). This fieldwork is often described as 'research in the wild' [22,23] and reflects the fact that one cannot fully grasp or appreciate the effects of technology without studying it in situ, as opposed the idealized setting of a laboratory. Arguably, the most challenging HCI experiments – even if not expressly done as such – are the deployments of new and disruptive technologies, normally outside the research setting (and thus any kind of research ethics process). For instance, Google Glass was launched onto the unsuspecting public, and attracted considerable opprobrium, notably including criminal and civil actions against its users [28,29] as well as risking the undesirable development of the law more generally [8,17]. In reality, how the legal system might react is a behavioral question: namely *how* and *why* do members of the judiciary act in response with being presented with a given technology? Knowing the answer to this question in advance is essential in the proper design of such technologies, especially where they might come to be depended upon by vulnerable minorities (e.g., as with Glass, as a potential assistive technology for people with disabilities).

To the knowledge of the author, there has been little in the way of HCI researchers conducting litigation *in person* and documenting it for the benefit of the academic community. At the very least, this has not been formalized as an approach. Courthouses appear to be a 'wild' where the HCI community has feared to tread. In some respects, this is surprising, given the multitude of challenging settings which HCI researchers have elected to personally conduct research and the increasing importance of legal (interaction) design [1,10,11,15]. The operation of the juridical process is of great importance for HCI: the law is a set of (sometimes ill-defined) constraints and opportunities that apply to the design and implementation of all information systems. Moreover, the legal system is intrinsically backwards looking [21] and only addresses cases or lawsuits brought by particular parties (in respect of events that have occurred some time ago), rather than the questions or arguments that are likely to be of relevance for the design of technologies that are in development [15]. Even in the cases that are in fact ruled upon, the reasoning is usually incomplete and focused upon what the judge considers the main points of the parties.

There is much to be learned that can only be found by *appearing* in the legal process: to design effective systems, we need to know how technologies will likely be reacted to so potentially harmful situations can be avoided. This means that there is a need to take an active part in the legal process in order to effectively understand how judicial decisions will be made in respect to future technologies. In effect, this requires research by conducting cases in person as a HCI researcher, which hereon in will be termed *research through litigation*. This paper develops and justifies the concept of research through litigation as a methodological innovation that is essential for legal design, and illustrates how this is practical, given that most cases take place in Tribunals (rather than traditional Courts).

## 2 RESEARCH THROUGH LITIGATION

### 2.1 What is Research through Litigation?

The notion of research through litigation is simple to state. It involves *deliberately* conducting legal proceedings with a view towards *understanding* the legal system and how it operates on the ground. After all, "*we cannot understand how a law works in practice simply by reading the law on pape*r" and laws commonly fail due to "*a lack of institutional capacity to implement the law*" [9]. This includes the operation of Courts and Tribunals.

Research through Litigation thus involves deliberately bringing proceedings in a Court or Tribunal, with a view towards testing how it operates. It is a form of 'mystery shopping', where the litigant does not disclose that they are there for research purposes, thus maintaining the Hawthorne Effect that is central to this activity [27].. The key quality is that it is generally done by non-lawyers participating as citizens in the legal process. The use of non-lawyers is essential, because otherwise the Hawthorne effect is unlikely to be maintained.

One advantage of litigation is the voluminous amount of documentation generated by the legal process. It frequently involves extensive and detailed correspondence, with submissions often being many thousands of words long. In some cases, it is possible to obtain extensive and detailed transcripts, as well as witness statements. The result is a rich corpus, with one proceeding alone often able to produce many hundreds of pages of documents. Those documents tell a story, namely the progression of a case and what happened within it, and thus provides a detailed record of what each actor said and did and for the most part, why they said they were doing so.

The resulting process is thus a hybrid of autoethnography and document analysis. As one litigates, one keeps track of the proceedings: each submission made is a judgment of what one thinks is going on and how to react to it. For practical purposes, these are auto-ethnographic observations (especially when considering the diverse approaches that can be taken towards autoethnography within HCI – see e.g. [6,13,24]). This is perhaps particularly so when they are accompanied by a variety of correspondence outside of the cases themselves. After all, I regularly correspond with journalists and various other actors in the FOI sphere, be they campaigners or litigants, or in some cases a combinations of the two – many of who would approach me for my advice (on one occasion I was even asked about an Assange case that was before the Tribunal, so that I could give background as to how it worked). Yet the record keeping has another advantage: it allows for what happened to be quoted at length, including from witness statements and transcripts, thus overcoming a common critique of the autoethnographic approach, because there is a solid corpus of documentation to analyze, the bulk of which was not generated by myself. It might be said that this is a form of automatic autoethnography, where the observations simply are created as part of the process.

## 2.2 How can Research through Litigation Contribute to Legal (Interaction)Design?

Traditional legal research has either been purely doctrinal in nature, or on occasions based on observations of Tribunals – it is not really focused on re-designing the legal system in most cases, but rather understanding it and criticizing it *within* the discipline of law [4]. Whilst it is true that one can observe some Tribunals, much of the legal system is not open to the public. Even where it is more open, potential attendees are rarely able to establish the issue in a given case or review the documents associated with the proceedings in advance. Neither do attendees decide *what* arguments are put, or *how* they are put across to a Court or Tribunal. Furthermore, when a journalist or academic visibly attends a Tribunal. one undoubtedly gets the Hawthorne effect [27], instead of observing the justice process when there is no expected scrutiny. Relying on 'natural experiments' is therefore insufficient to effectively probe the legal system. It prevents the asking of 'what if' questions, such as how a given technology will impact proceedings, how a Tribunal will react to particular submissions or how a Tribunal would treat individual litigants in particular circumstances. It also removes perspective: the truth is that there is only limited information obtainable from the written documentation of proceedings (in most cases, at best only the final judgment is available, rather than any transcript of what was said, or the arguments advanced by each party). This means that just reading the publicly available documentation itself is likely to lead to a misinformed reader. In other words, being an active part of the process is necessary to understand it.

The primary learnings from research through litigation is thus *information* on human behavior and preferences, which (crucially) could not be obtained by other means. However, it is also possible to shape the system by arguing 'test cases' before it: after all, design is a response to a set of circumstances and sometimes changing the circumstances, policies and rules removes constraints and therefore enables more ambitious designs. Just to give some examples, my own cases led to several changes and clarifications to the law itself – for instance, changing FOI law provides opportunities for creating new digital tools that assist requesters, whilst changing Tribunal practice can make cases more accessible for those who take part in them. It is also an example of how design can contribute to the law: after all, an argument on what the law ought to be can often be derived from how the system should be designed to work more effectively.

## 3 LEARNINGS FROM RESEARCH THROIUGH LITIGATION

### 3.1 My 'research through litigation'

Over a seven-year period, I have participated in a number of Freedom of Information Act (2000) proceedings in the United Kingdom. This arose out of being told that it took too long to find documents, for reasons that I knew were plainly wrong as a matter of computer science (and which turned out to be systematic issue [14]). The effect of this was that I ended up arguing and challenging fundamental issues of computer science and law, which tested them in such a way that no other person had considered them. I should add this was not the plan, it was just a case of one thing leading to another.

### 3.2 What sort of things did I end up learning about and evidencing?

There is a wide range of examples that I encountered in my somewhat strange trip through the Tribunal system. These include some remarkable occasions, which are detailed at length in my forthcoming paper [16]. The below gives just a small flavor of the types of things one can learn from this process.

- **Example 1: Closure of the Information Tribunal due to difficulties in organising PDF's.** The (Information) Tribunal jurisdiction was closed down because the UK Information Rights regulator found it too difficult to organise PDFs into bundles. The then President of the relevant Tribunal accepted this assertion. When this matter was raised in the press, the Judicial Press Office wrote to the journalist threatening to sue the journalist 'on behalf of the judiciary', on the apparent instructions of the then President.
- **Example 2: Using Microsoft Excel requires an 'academic' computer scientist.** A senior Judge responsible for Information Rights asserted that using relatively simple procedures in Microsoft Excel was something that required the "*technical competence of a computer science academic, and not one that may reasonably be expected of a public authority*". The Judge then attempted to prevent me from obtaining a transcript of the hearing before her. I successfully appealed that decision to the Upper Tribunal with the result of confirming in law the right of litigants to obtain transcripts of hearings (with the case being reported as Kirkham v Information Commissioner (GIA) (Tribunal procedure and practice - record of proceedings) [2019] UKUT 381 (AAC)). An account of the worrying situation before the First-tier Tribunal as covered by a member of the press can be found here (https://www.computerweekly.com/news/252509483/Government-bodies-refuse-FOI-requests-on-basis-of-misleading-database-search-times-says-academic)
- **Example 3: Trying to do remote hearings via ISDN.** Before the pandemic, a Tribunal attempted to force me to use ISDN at the cost of around $1000 an hour to connect to hearings. In another case, a different Tribunal struck it out for my moving abroad, on the basis it was too challenging to do remote hearings. This strike out was overturned by an exasperated judge of the Upper Tribunal. The issue of access to online justice using an appropriate system was finally resolved by Upper Tribunal Judge Thomas Church, who decided to make directions for the courts to be configured in line with my instructions.
- **Example 4: Requiring in person hearings instead of using Zoom.** In a hearing, it was asserted that claimants should not do hearings from abroad but wait until they returned. This was said to be a practice from "*War Pensions*" cases (where many applicants to the tribunal lived abroad and have significant disabilities). In effect, this group would have had to wait an extended time of their pensions due to the Tribunals' refusal to do remote hearings. It was obvious and most worrying disability discrimination.
- **Example 5: Inexpert Judicial 'Data Protection'.** This example is particularly worrying. The Judge responsible for tribunals data protection (i.e. the "*judicial data protection network lead*"[1] responsible for "*supervis[ing] the data processing activities* of Tribunals") expressed the belief that "*you do not build your systems around the rogue, do you?*". Of course, computer security is fundamentally about protecting from rogue actors.

I also changed the law on multiple occasions, including where Judge Jacobs reformed the right of access under the Freedom of Information Act (2000) to be far more favourable to FOI requesters, and one occasion where Judge Markus QC decided to create a right to transcripts in the First-tier Tribunal.

### 3.3 How does this contribute to Legal (Interaction) Design

Legal design has only been subject to limited work within HCI and tends to be at the boundaries of the legal system – for instance testing how laws operate on the ground [2,14,18], rather than how technologies are evaluated and interpreted by courts and tribunals, or the occasional deployment of a system that is designed to help (e.g [26]). Whilst there have been occasional efforts to develop and explore design principles for the legal system (see e.g. [3,12,19,25]), these efforts are at best nascent. This is a space where a lot of work needs to be done, with legal (interaction) design to date being criticized for "*its lack of theoretical and methodological rigour [by] from the design and legal innovation communities*" [3]. One issue is a lack of connectivity between law and HCI on the ground, with there being numerous legal technology projects being implemented that simply overlook the key principles from HCI (e.g. online court systems at the costs of hundreds of millions that Judges detest so much they attempt to use Zoom instead [30]).

---

[1] https://www.judiciary.uk/wp-content/uploads/2019/01/Supplementary-SPT-report-Dec-2018_final.pdf

Against that context, Research through Litigation allows us to:
- Put pressure on the legal system to change and even obtain new case law that is more favorable to the design of technologies.
- Understand judicial attitudes, so that we can understand how new technologies might be received and/or addressed.
- Document and understand existing procedural and practical barriers, so we can either tackle them, or work around them.

In short, research though litigation is a novel approach towards advancing Legal (Interaction) Design.

## 4 CONCLUSION

This article has developed an account of how the process of litigation can be adopted as a form of (legal) design research. Drawing on the setting of the Information Tribunal, this approach was validated using case studies of actual litigation conducted by the author. This new approach could be fruitful for minimizing the risk of challenging technologies (and therefore their users) falling foul of the legal system, thus ensuring that future technologies are more harmoniously integrated into society going forwards. At the same time, this represents an innovative new direction for legal design work, as well as illustrating the advantage of less formal methods for implementing design in this (highly formal) space.